# Understanding of Normal and Abnormal Hearts by Phase Space Analysis and Convolutional Neural Networks

Bekir Yavuz Koc[1], Taner Arsan[1,*], Önder Pekcan[2]

[1]*Department of Computer Engineering, Kadir Has University, Istanbul, Turkey*
[2]*Faculty of Engineering and Natural Sciences, Kadir Has University, Istanbul, Turkey*
*Corresponding Author: arsan@khas.edu.tr

**Abstract**

Cardiac diseases are one of the leading mortality factors in modern, industrialized societies, which cause high expenses in public health systems. Due to high costs, developing analytical methods to improve cardiac diagnostics is essential. The heart's electric activity was first modeled using a set of nonlinear differential equations. Following this, variations of cardiac spectra originating from deterministic dynamics are investigated. Analyzing a normal human heart's power spectra offers His-Purkinje network, which possesses a fractal-like structure. Phase space trajectories are extracted from the time series electrocardiogram (ECG) graph with third-order derivate Taylor Series. Here in this study, phase space analysis and Convolutional Neural Networks (CNNs) method are applied to 44 records via the MIT-BIH database recorded with MLII. In order to increase accuracy, a straight line is drawn between the highest Q-R distance in the phase space images of the records. Binary CNN classification is used to determine healthy or unhealthy hearts. With a 90.90% accuracy rate, this model could classify records according to their heart status.

## 1. Introduction

It is well known that the human heart has an electrical activity that can be detected by measuring the potential difference from various points on the body. Electrocardiogram (ECG), which measures electric potential versus time, has three distinct parts. P-wave is associated with the excitation of the atria, the QRS complex presents the ventricles (His-Purkinje network), and T-wave shows the recovery of the initial electrical state of ventricles (Figure 1a). Although ECG presents periodic behavior, some irregularities can be observed in the details of the records. In fact, these irregularities belong to the intrinsic part of heart activity and/or to the random noise that can be found in such systems. These activities in ECG spectra help us to understand cardiac dynamics. Cardiac oscillations are sometimes perturbed by unpredictable contributions, which are part of the cardiac dynamics and, therefore, physiologically important. Due to these reasons, the heart is not a perfect oscillator and/or cardiac muscles do not continuously vibrate harmonically.

For researchers, transforming a sequence of values in time to a geometrical object in space is a highly considered topic [1]. This procedure replaces an analysis over time with an analysis in space. Space is considered as phase space and the transforming of time series into an object called an embedding (Figure 1b). After this step, topological properties of the object are determined based on its fractal dimension. In this fractal dimension, not the original time series but the fractal dimension characterizes the properties of the phase space set. In the phase space set, the measured dimension determines whether a data set is generated by random or deterministic processes. A large fractal dimension value indicates the time series' random generation. The number of variables and equations is immense, and there are no ways to predict the future values from the early parameters. This shows that the multiplicity of interacting factors precludes the possibility of understanding how the underlying mechanisms work.

On the other hand, a low fractal dimension value indicates that the data is generated by deterministic mechanisms based on a small number of independent variables, which helps to understand how the values in the past can be used to predict the values in the future. The system would be considered chaotic if the mechanism generating the data is deterministic, but the time series behaves like generated by random processes. A chaotic system is deterministic but not predictable in the long range. In a deterministic approach that is not chaotic, the value of a variable at a given time can be used to always generate the value of that variable in the future.

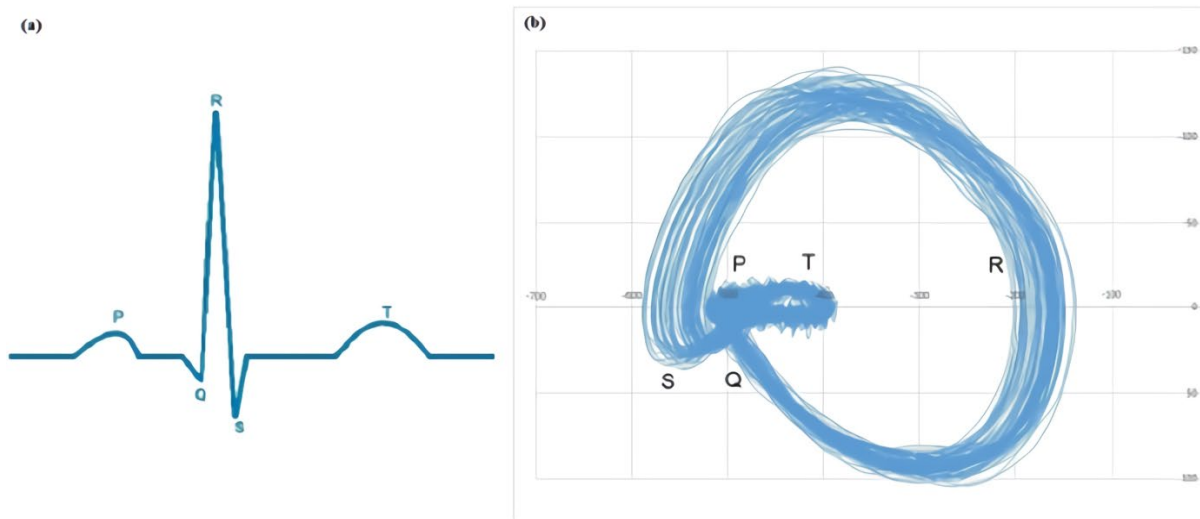

Figure 1. PQRST in ECG and phase space. (a) ECG (b) Phase space.

At early times, the heart's electric activity was modeled using a set of nonlinear differential equations [2-4]. Afterward, variations of cardiac spectra originating from deterministic dynamics called "chaotic dynamics" highly sensitive to the heart's initial conditions are analyzed [5-6]. Analyzing the power spectra of a normal human heart shows that the His-Purkinje network possesses a fractal-like structure [7], and the existence of chaotic behavior of the heart can be found from the phase space picture by evaluating a fractal dimension, $D$, where phase space trajectories are extracted from the time series graph of ECG [8]. Lower values of $D$ indicate more coherent dynamics. If $D = 1$, the oscillation is periodic, and the phase space picture shows a limiting cycle. However, $D$ becomes larger than one when a limiting cycle is perturbed by random noise [9]. $D$ has non-integer values greater than two when the system becomes chaotic or strange attractor [9]. In this case, although trajectories in time do not converge towards a limiting cycle, they stay in a bounded region in the phase space.

It is also shown that a short-term heart rate variability analysis yields a prognostic value in risk stratification, independent of clinical and functional variables [10]. However, the detailed description and classification of dynamical changes using time and frequency measures are often insufficient, especially in dynamical diseases characterized by Mackey and Glass [11-12].

In [13], Niels et al. try to find a solution for the normal heart rate being chaotic due to the respiration question. In their work, they give an example of the sequence of respiration on heartbeat dynamics, showing that respiratory modulations of heart rate and blood pressure can mostly explain observed fluctuations. Recently, the development of a fast and robust method that can be applied to multichannel physiologic signals has been reported [14]. This method elaborates on removing a selected interfering signal or separating signals arising from temporally correlated and spatially distributed signals such as maternal or fetal ECG spectra. Convolutional neural networks (CNNs) method was also applied to patient-specific ECG classification for real-time heart monitoring [15].

In "Heartbeat classification fusing temporal and morphological information of ECGs via ensemble of classifiers" study published in 2019 by Mondéjar-Guerra, ECG-based automatic classifications are combined with multi-ended Support Vector Machine Models (SVMs). This methodology is based on the time interval of subsequent beats and beats morphology in electrocardiogram characterization. Different descriptors based on wavelets, local binary patterns (LBP), and high-order statistics (HOS) values were used in the study. Rather than the concatenating process of all features in a single support vector machine model, these features are used differently. They are used as input to teach the different support vector machine models for various features. Before getting the ultimate classifying result, these models are joined with the sum, majority, and product methodologies. In the study, the Daubechies wavelet function (db1) is used, and a twenty-three-dimensional descriptor with three decomposition levels is built. With the division of each beat in five intervals, a ten-dimensional attribute is created to use in HOS (High Order Statistics). With LBP, which includes eight neighbor uniform codes, fifty-nine-dimensional descriptors are used with 1D-LBP.

Additionally, a morphological descriptor is implemented in the study. It depends on the sample and amplitude distance by calculating Euclidean distance in the R-peak and the other four waves of the beat. The study reported that the best parameters were adjusted, and the models were retrained with a training set and tested with a testing set. It is specified that due to the unbalanced dataset, the mean or overall accuracy did not show the true performance of the classifier. On top of that, it is indicated that if the classifier is assigned to a normal testing class, it is expected that it would classify with at least eight-nine percent overall accuracy [16].

Nowadays, it is well understood that cardiac diseases are one of the main reasons of mortality in modern, industrialized societies, and they cause high expenses in public health systems. Due to high costs, it is essential to develop analytical methods to improve cardiac diagnostics. In this work, we investigated 44 ECG recordings taken from abnormal and normal hearts, and CNN was applied for binary classification. We aim to classify healthy and unhealthy hearts by converting the ECG recordings into phase spaces in the third-order derivate Taylor series.

The rest of the paper is organized as follows. Section 2 introduces the algorithms and describes the dataset used in this study. Results obtained from phase space diagrams of healthy and unhealthy records are given and discussed in Section 3. The conclusions, Section 4, complete our paper.

## 2. Methodology
### 2.1. Phase Space Approach

We construct the phase space based on heart voltage values over time (i.e., $f(t)$) and their first derivative (i.e., $\frac{df(t)}{dt} = f'(t)$) (see Figure 1b). We know the function $f(t)$, but we need to obtain $f'(t)$. When we sketch the diagram of $f(t)$ and $f'(t)$, we can construct the phase space graph. As described in [17-19], the simple approximation of $f'(x)$, the first derivative of a function $f$ at a point $x$ is specified as the limit of a difference quotient as follows:

$$f'(x) = \lim_{h \to 0} \frac{f(x+h) - f(x)}{h} \tag{1}$$

If $h > 0$, meaning that h is a finite positive number, then

$$f'(x) = \frac{f(x+h) - f(x)}{h} \tag{2}$$

is called the first-order forward difference approximation of $f'(x)$. The approximation of $f'(x)$ can be obtained by combining Taylor series expansions. If the values $(x_{i+1}, f_{i+1})$, $(x_{i+2}, f_{i+2})$ and $(x_{i+3}, f_{i+3})$ are known, the first-order derivative of $f_i(x)$ can be calculated as shown by $f'_i(x)$.

$$f_{i+1} = f_i + \frac{f_i'}{1!}h + \frac{f_i''}{2!}h^2 + \frac{f_i'''}{3!}h^3 + \cdots + \frac{f_i^n}{1!}h^n + R_n \tag{3}$$

$$(x_{i+1}, f_{i+1}) \to f_{i+1} = f_i + \frac{f_i'}{1!}h + \frac{f_i''}{2!}h^2 \to f_{i+1} = f_i + h f_i' + \frac{h^2}{2!}f_i'' \tag{4}$$

$$(x_{i+2}, f_{i+2}) \to f_{i+2} = f_i + \frac{f_i'}{1!}2h + \frac{f_i''}{2!}(2h)^2 \to f_{i+2} = f_i + 2h f_i' + 2h^2 f_i'' \tag{5}$$

$$(x_{i+3}, f_{i+3}) \to f_{i+3} = f_i + \frac{f_i'}{1!}3h + \frac{f_i''}{2!}(3h)^2 \to f_{i+3} = f_i + 3h f_i' + \frac{9}{2}h^2 f_i'' \tag{6}$$

The second derivative $f_i''$ is canceled by adding Equations (4), (5) and (6) after multiplying them by 18, -9 and 2, respectively. After this step we obtain:

$$18 f_{i+1} - 9 f_{i+2} + 2 f_{i+3} = 11 fi + 6 h f_i' \tag{7}$$

Finally, third-order forward difference Taylor series approximation of $f'(x)$ is obtained as follows:

$$f_i' = \frac{1}{6h}(-11 f_i + 18 f_{i+1} - 9 f_{i+2} + 2 f_{i+3}) \tag{8}$$

When we know the first three values of the signal, we can easily obtain the first derivative and we can sketch the phase space of this signal.

## 2.2. Convolutional Neural Networks

In order to make classifications in the field of computer vision, Convolutional Neural Networks (CNN), which are inspired by human visual perceptions, have been discovered [20]. As shown in Figure 2, CNN can have one or more layers and increases its complexity with high-level features.

CNN has weights coefficients that provide learning which is equivalent to classical neural networks. CNN has achieved great success, especially with problems with images. There are convolutional layers, a pooling layer, fully connected layers, and a training process in CNN [21].

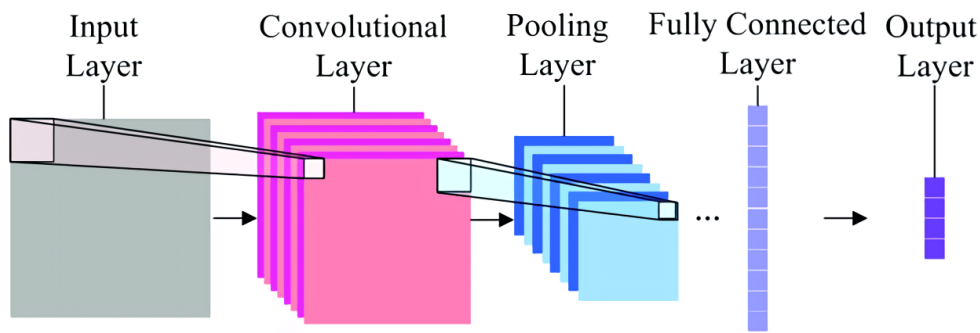

Figure 2. General Structure of Convolutional Neural Networks [22].

### 2.2.1. Convolutional Layer

In the convolution layer, image or video information is extracted from the input data, as shown in Figure 3. Convolution works by learning image attributes and considering or preserving spatial relationships between pixels. If each image is considered a pixel matrix, convolution reduces data blocks to a smaller size than the given matrix. By moving the "feature detector" or "filter" matrix, the "Feature Map" or "Evolved Feature" matrix can be obtained (see Figure 3). This sequence is repeated when the input image is transformed into multiple feature map sets [23].

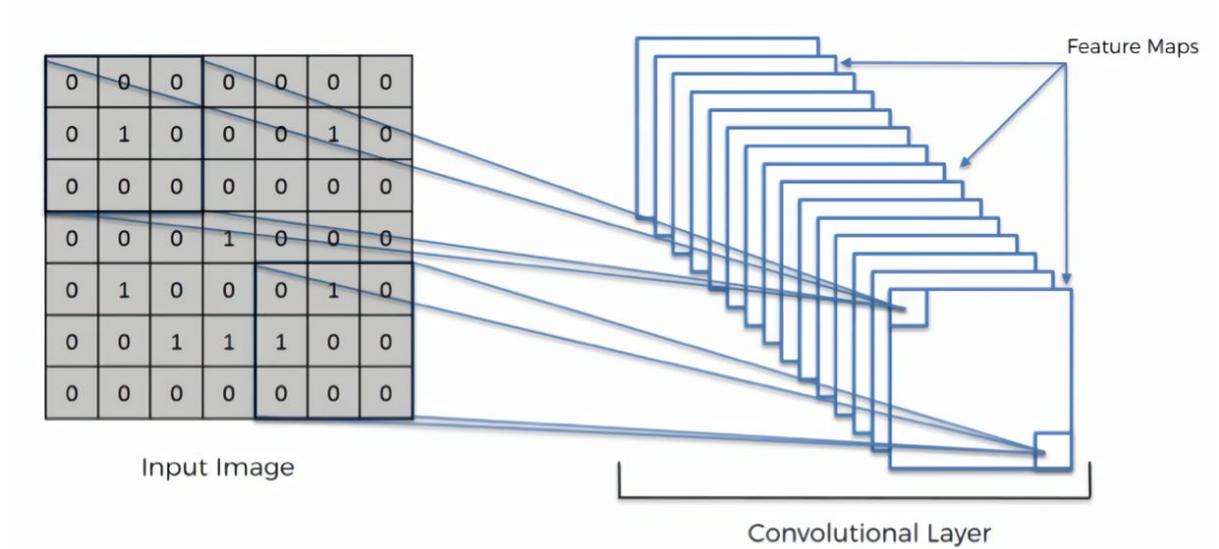

Figure 3. Feature map in the convolution layer [24].

### 2.2.2. Activation Functions and ReLU

In CNN, the activation function is frequently used as Rectified Linear Unit (ReLU). This is the stage after the convolution process. ReLU is used for including nonlinearity in training. ReLU produces zero if a function gets a negative value, or it returns the positive value, as in Figure 4.

$$f(x) = \max(0, x) \qquad (9)$$

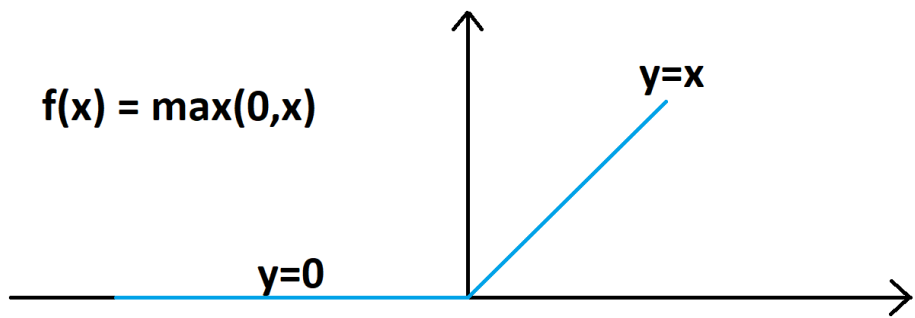

Figure 4. Rectified Linear Unit (ReLU).

### 2.2.3. Pooling Layer

When there is a lot of input data, the number of trainable records and the computational load can be redundant. This load can be reduced by reducing the width and height and keeping the input data's depth. Moreover, with the pooling layer, reducing the size provides information loss and prevents the model from memorizing [25]. Mostly, max pooling is used in the industry. Max pooling is used by taking the highest, which is the maximum value, in each specified part of the attribute map [26].

$$S_i = \max a_i \ (i \in R_j) \tag{10}$$

### 2.2.4. Fully Connected Layers

Fully Connected Layers are the last layer of the CNN model. In order to transform the CNN model into a classical artificial network, a flattening process needs to be used. The flattening process indicates that the concatenation and the final (or any) output of the convolutional neural network, i.e., a 3-dimensional matrix, the flattened output of the convolutional layer or final merging is transmitted as input to the fully connected layer that will be transformed into a vector as shown in Figure 5. Softmax or sigmoid activation functions are used in fully connected layers in classification determination. These functions will provide a non-negative value up to one in the output layer [27].

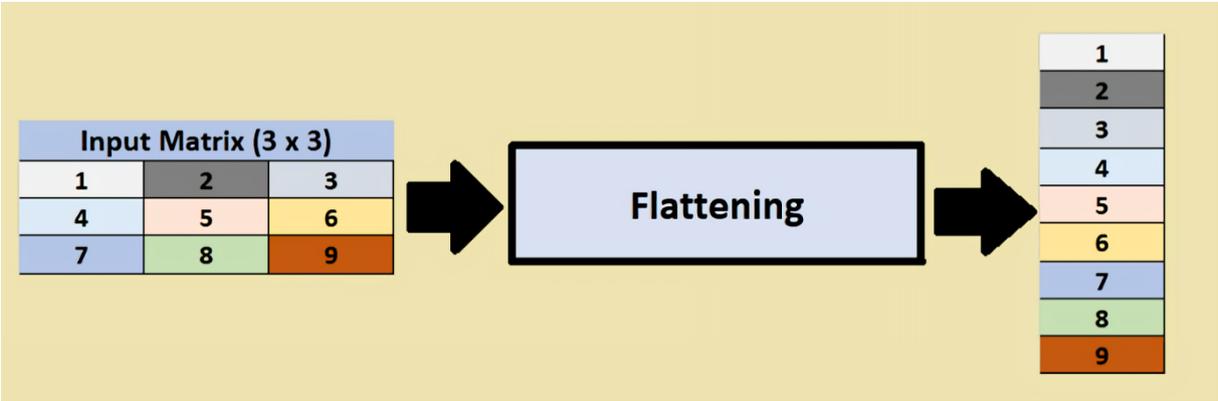

Figure 5. Flattening in CNN.

### 2.3. Model Training and Loss Function

CNN aims to minimize the wrong classes in the test set. In order to achieve that, the weight coefficients are updated. Since the correct tags of training examples are known during the training of the network, the difference between the tags estimated by CNN (with the current weight coefficients) and the actual class tags defined in the training set is calculated with the loss function.

The parameters of neurons in convolutional and fully connected layers are updated with a gradient descent algorithm. With this algorithm, the class scores of the convolutional network can be calculated according to the tags in the training set for each image [21].

$$w_{ij}^{t+1} = w_{ij}^{t} - \alpha \, (dL/\, dw_{ij}) \quad (11)$$

where $w_{ij}$ is the weight coefficient of the $j$ neuron in layer $i$, $t$ is the iteration step, $\alpha$ is the learning rate, and $L$ is the loss function [28].

### 2.4. Dataset

MIT-BIH Database contains two-channel half-hour 48 extracts of ambulatory ECG recordings taken from 47 people by the BIH Arrhythmia Laboratory between 1975 and 1979. Each of the 48 recordings is nearly over 30 minutes. In 46 of the 48 recordings, the subjects were measured with the MLII signal obtained by placing the electrodes on the subject's chest. Record 102 and record 104 could not measure in the MLII signal; they are measured in the V5 signal. These two records are excluded from the database in this thesis due to this reason. Figure 6 presents the records for healthy (101) and unhealthy (210) people taken from MIT-BIH Database. In addition, the records with paced beats were not considered, which are record 107 and record 217.

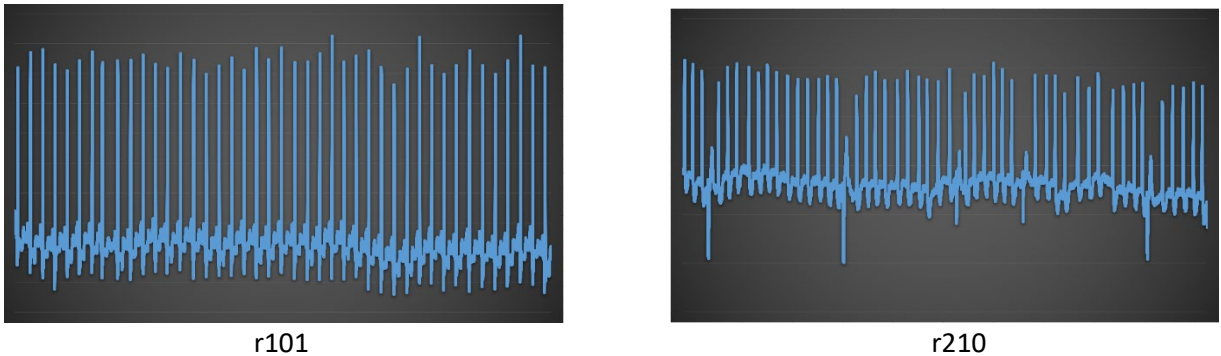

Figure 6. ECG signal of two different records. r101 Healthy, r210 Unhealthy.

In the "Heartbeat classification fusing temporal and morphological information of ECGs via ensemble of classifiers" study which was published in 2019 by Mondéjar-Guerra, ECG-based automatic classifications are combined with multi-ended SVMs. The database contains nearly 110.000 beats in total. The beats are grouped under SVEB (Supraventricular ectopic beat), VEB (Ventricular ectopic beat), F (Fusion), normal and unknown. On the other hand, in this study, records data are taken individually, and CNN methodology is applied for 47 records. In order to increase the success rate of the classification, a single line was drawn in the phase space images from the Q wave connected to the highest R wave.

## 3. Results and Discussions

Phase Space Diagrams of Healthy and Unhealthy records are presented in Figure 7 and Figure 8, respectively, for the CNN analysis, where Keras and Tensorflow libraries are used in Python to implement the CNN model.

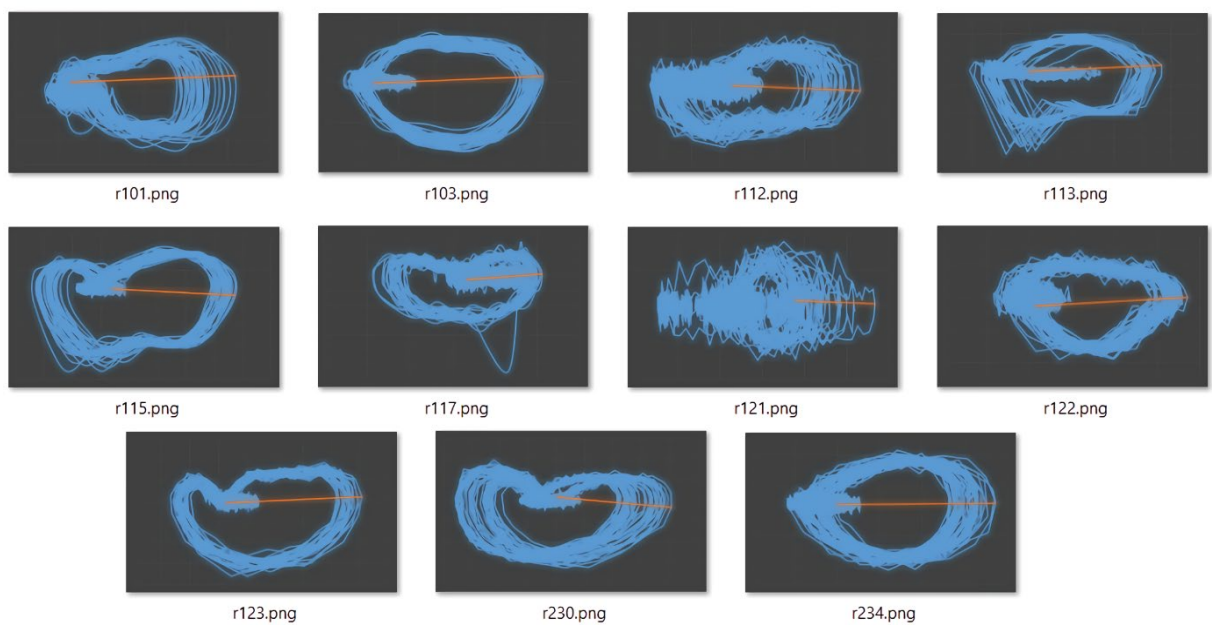

Figure 7. Phase Space Diagram of Healthy Records:
Healthy Records, Phase Space Diagram: 101, 103, 112, 113, 115, 117, 121, 122, 123, 230, and 234.

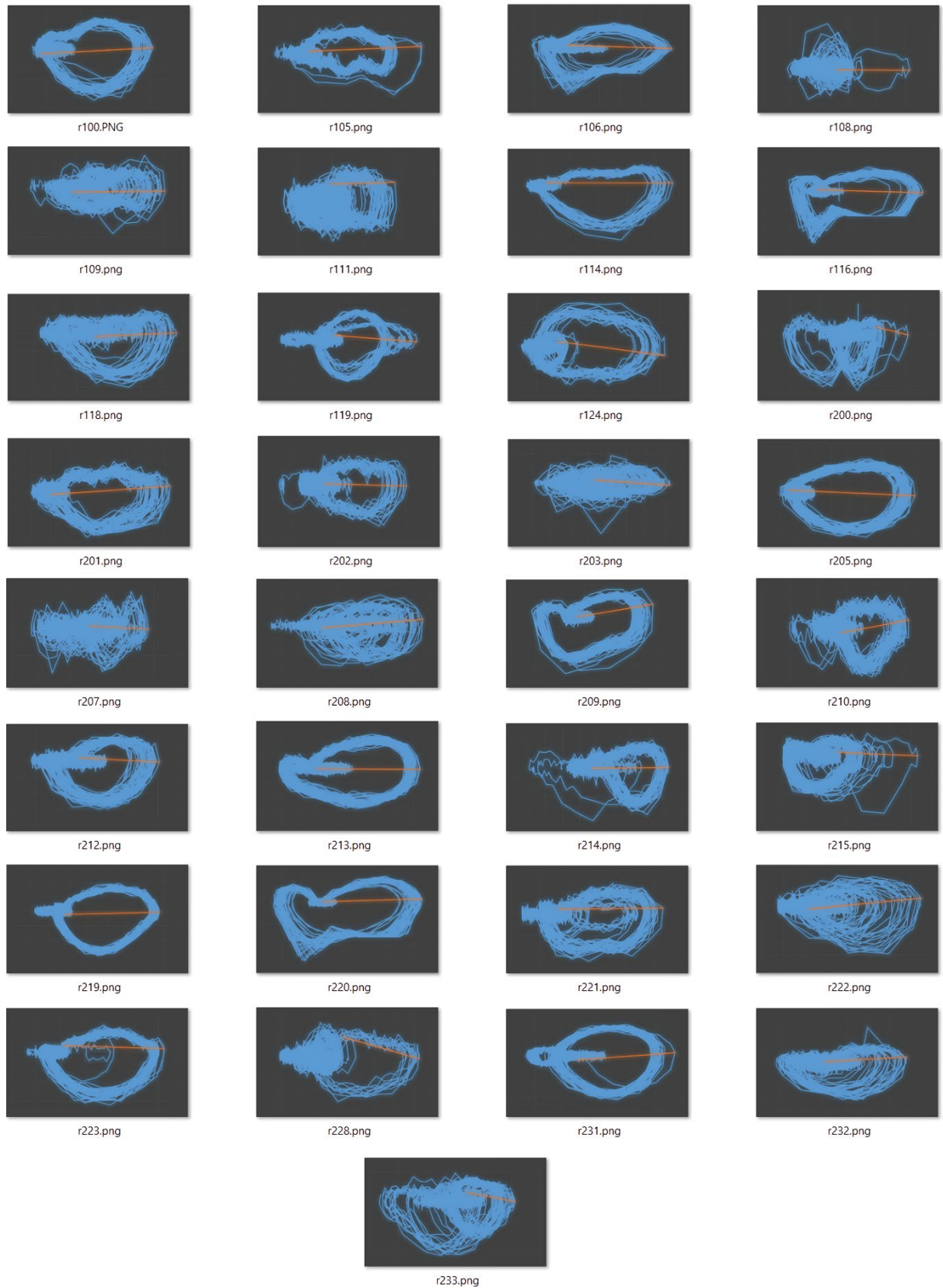

Figure 8. Phase Space Diagram of Unhealthy Records:
Unhealthy Records, Phase Space Diagram: 100, 105, 106, 108, 109, 111, 114, 116, 118, 119, 124, 200, 201, 202, 203, 205, 207, 208, 209, 210, 212, 213, 214, 215, 219, 220, 221, 222, 223, 228, 231, 232 and 233.

Input shape of 64 x 64 x 3 feature vectors were fed to the convolution layers followed by the pooling layer, where the padding features were kept the same as the input to the pooling layer. Dense layers are used to minimize the size of the output vector gradually.

The data augmentation method is used to create more pictures per epoch (see Figure 9). Zoom range, shear range, and horizontal flip parameters are adjusted with the Keras library. Two convolutional layers and max pooling are used related to the ReLU activation function. One hundred twenty-eight neurons are used in the hidden layer, and one neuron is used for the output layer due to binary CNN classification. One hundred seventy-five epochs are used to train the model.

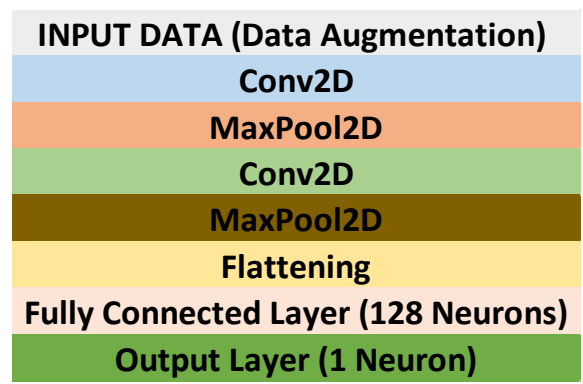

Figure 9. The architecture of CNN model.

The training set includes 33 records, and the test set consists of 11 records (Table 1). The training set ratio is 75%, and the test set ratio is 25%. The Healthy test set includes three different records, and the training set contains eight different records. 27.27% of healthy records exist in the testing set and 72.72% in the training set. The disease test set includes eight disease records, and the training set contains 25. There 24.25% of disease records exist in the test, and 75.75% exist in the training set. The model generally results in higher accuracy in training and testing sets. Although there is an outlier epoch performance in the $105^{th}$ epoch after the $75^{th}$ epoch, the training set accuracy and testing set accuracy increase with higher parallel performance (Figure 10).

Table 1. Training Set and Test Set

| SET | Training Set | Testing Set |
|---|---|---|
| Healthy records | 101, 113, 115, 117, 121, 122, 123, 230 | 103, 112, 234 |
| Unhealthy records | 106, 108, 109, 114, 116, 118, 119, 124, 201, 203, 205, 207, 208, 209, 214, 215, 219, 220, 221, 222, 223, 228, 231, 232, 233 | 100, 105, 111, 200, 202, 210, 212, 213 |

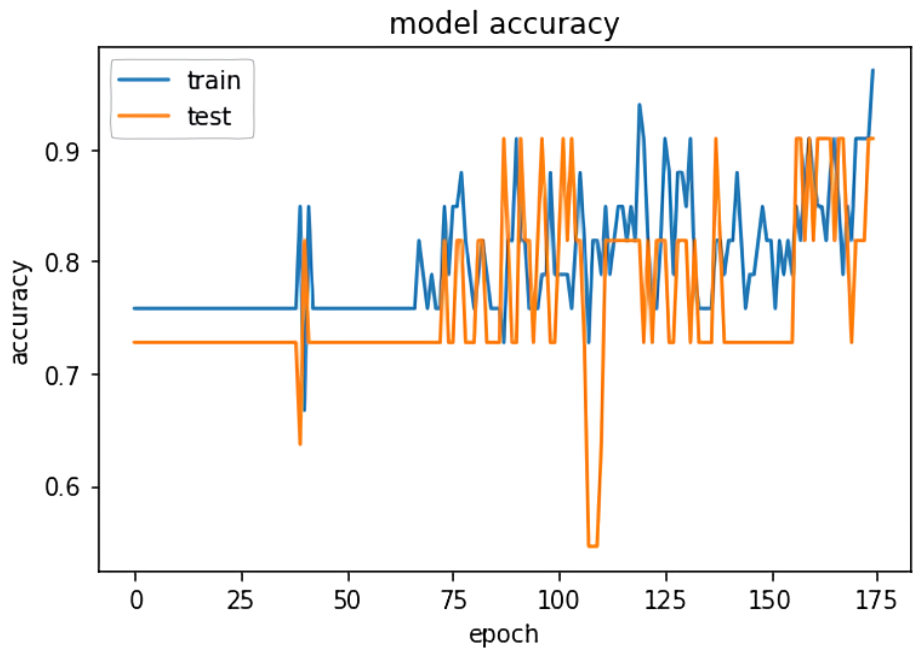

Figure 10. Phase Space model loss comparison of training set and testing set in altered dataset.

It can be seen that the training and testing loss are parallel. The results show that the training ratio and data augmentation prevented the overfitting problem (see Figure 11). Therefore, the model can train itself to classify the training dataset and accurately classify the test dataset. As a result, model accuracy improves in the training and test sets with a well-trained model to classify healthy hearts and diseased hearts.

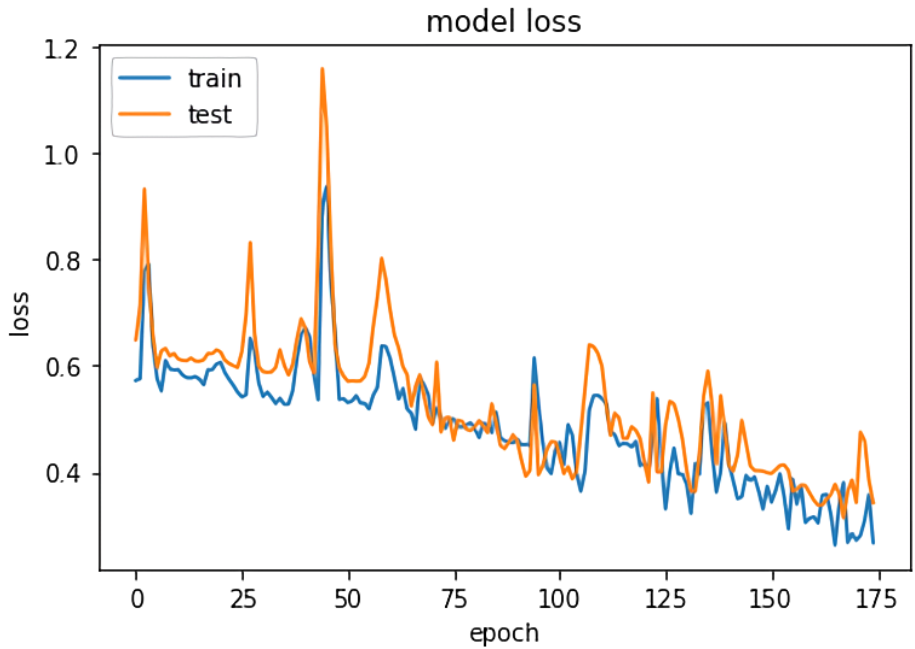

Figure 11. Phase Space accuracy comparison of training set and testing set in altered dataset

Among 11 test records, the model can correctly classify 10 test records with only one faulty record. Healthy test set accuracy is 100%. Among three healthy records, the model can classify all the records correctly. On the other hand, the disease test set accuracy is 87.5%. The model can correctly classify all records except r100 among eight disease records. This record is classified under a healthy record, although it is an unhealthy heart. Figure 12 summarizes the output of the phase space results with altered data records.

| Outputs of Phase Space Results | | | | | | | |
|---|---|---|---|---|---|---|---|
| Epoch Count | | Accuracy | | Val Accuracy | | | |
| 175 | | 96.97% | | 90.90% | | | |
| | | Healthy | | | | | |
| | r103 | | r112 | | r234 | | |
| | Healthy | | Healthy | | Healthy | | |
| Disease | | | | | | | |
| r100 | r105 | r111 | r200 | r202 | r210 | r212 | r213 |
| Healthy | Disease | Disease | Disease | Disease | Disease | Disease | Disease |

Figure 12. Phase Space model outputs of training set and testing set in altered dataset

As shown in Table 2, it can be observed that model can classify health records with the best accuracy reaching almost 100% in the test set. The model has 87.5% accuracy that classifies disease records. There are no overfitting problems reported in this model training. The training accuracy and testing accuracy are similar, with very high accuracy rates.

Table 2. Results of the Model

| Model | Accuracy |
|---|---|
| Training Set | 96.97% |
| Test Set | 90.90% |
| Healthy Test Set | 100.00% |
| Disease Test Set | 87.50% |

## 4. Conclusion

After the transformation of the data into phase diagrams, despite the insufficient number of data, this model can achieve remarkable accuracy rates. The model can classify healthy and unhealthy heart image records with 90.90% accuracy using only 44 images and the Data Image Generator feature.

Overfitting problems are inevitable when the number of data is limited. CNN can extract essential features and make valuable distinctions between healthy and unhealthy hearts as a result of transforming the data into phase diagrams. The overfitting problem is outperformed, and the model obtains a healthy learning process.

The model holds an excellent performance for the determination of the classification of healthy hearts. The three records (r103, r112, and r234) are classified correctly. Furthermore, the model learned the essential features of healthy hearts and clearly classified unhealthy hearts. The results show the model's success with 87.50% accuracy rates in the classification performance of unhealthy hearts.

Finally, it may be concluded that more than 2 layers in convolutional layer can be included to improve the model design and increase the accuracy. In addition, more hidden layers can be added to update weights to classify the outcome.